\documentclass[12pt]{article}

\usepackage{braket}
\usepackage{mathrsfs}
 \usepackage{extpfeil}
\usepackage[nosort]{cite}
 \usepackage{graphics}
\usepackage{epstopdf}

\numberwithin{equation}{section}
\begin{document}

\begin{center}
{\LARGE{\bf{On some properties of number-phase Wigner function }}}
\end{center}

\bigskip\bigskip

\begin{center}
Maciej Przanowski\footnote{E-mail address:  maciej.przanowski@p.lodz.pl} and Przemys\l aw Brzykcy\footnote{E-mail address:  800289@edu.p.lodz.pl}  
\end{center}

\begin{center}

{\sl  Institute of Physics, Lodz University of Technology,\\ W\'{o}lcza\'{n}ska 219, 90-924 \L\'{o}d\'{z}, Poland.}\\
\medskip

\end{center}

\vskip 1.5cm
\centerline{\today}
\vskip 1.5cm

\begin{abstract}
It is shown that the number-phase Wigner function defines uniquely  the respective density operator.
Relations between the Glauber-Sudarshan distribution $\mathcal{P}(\alpha)$ and 
the number-phase Wigner function is found. This result is then generalised to 
the case of the Cahil-Glauber distributions $\mathcal{W}^{(s)}(\alpha)$, $-1\leq s \leq 
1$.
\end{abstract}



 \section{Introduction} \label{sec1} 
 In recent works \cite{1,2,3} we have developed a theory of quantum phase 
 which is based on some enlarging of the Fock space to the Hilbert space of the 
 square integrable functions on the circle, $L^2(S^1)$.
 Then the well known machinery of the Naimark projection is employed to find the 
 respective objects in the original Fock space. Such an approach to quantum 
 phase have been considered by many authors ( for references see \cite{1,2}). In 
 our case we were able to define also a number-phase quasiprobability 
 distribution which we called the \textit{number-phase Wigner function} 
 \cite{2,3}.
 In short our construction can be introduced as follows:
 Firstly, define the self-adjoint operator $\hat{\Omega}(\phi,n)$
 \begin{equation}\label{eq1.1}
   \hat{\Omega}(\phi,n):= \pi \left\{ \ket{n} \braket{n|\phi } \bra{\phi}  + \ket{\phi} \braket{\phi|n } \bra{n}  \right\}
 \end{equation}
 where $\ket{n}$ is a normalised eigenvector of the number operator $\hat{n}$ i.  e.
 \begin{equation}\label{eq1.2}
   \hat{n}\ket{n}=n \ket{n}, \quad \braket{n'|n}=\delta_{n'n}\,, \quad 
   n,n'=0,1,2,\dots
 \end{equation}
 and $\ket{\phi}$ stands for  the phase state vector
 \begin{equation}\label{eq1.3}
   \ket{\phi} = \frac{1}{\sqrt{2\pi}}\sum_{n=0}^{\infty} e^{in\phi}\ket{n}
 \end{equation}
 Then the \textit{number-phase Wigner function} is defined as
 \begin{equation}\label{eq1.4}
   \varrho_{W}(\phi,n) := \frac{1}{2\pi} \mathrm{Tr} \left\{ \hat{\varrho}\, \hat{\Omega}(\phi,n) \right\} 
   = \mathrm{Re}\left\{  \braket{\phi| \hat{\varrho} | n} \braket{n|\phi}  \right\}
 \end{equation}
 where $\hat{\varrho}$ is the density operator of a given quantum system.
 The operator $\hat{\Omega}(\phi,n)$ given by ($\ref{eq1.1}$) can serve as a 
 \textit{number-phase Stratonovich-Weyl quantizer}. Namely, for any classical 
 number-phase function $f=f(\phi,n)$ one can assign the respective operator $\hat{f}$ 
 according to the rule (the \textit{generalized Weyl quantization})
 \begin{equation}\label{eq1.5}
   \hat{f} := \frac{1}{2\pi}\sum_{n=0}^{\infty} \int_{-\pi}^{\pi} f(\phi,n) \hat{\Omega}(\phi,n)
   \mathrm{d}\phi.
 \end{equation}
 Employing the well known formulae
 \begin{equation}\label{eq1.6}
   \sum_{n=0}^{\infty} \ket{n}\bra{n}=1, \quad \ \int_{-\infty}^{\infty} 
   \ket{\phi}\bra{\phi}
   \mathrm{d}\phi=1
 \end{equation}
 we quickly find that if $f=f(n)$
 \begin{equation}\label{eq1.7}
   \hat{f}= f(\hat{n})
 \end{equation}
 and if $f=f(\phi)$ then
  \begin{equation}\label{eq1.8}
   \hat{f}= \int_{-\pi}^{\pi}  f(\phi)\ket{\phi}\bra{\phi}
   \mathrm{d}\phi.
 \end{equation}
 For any $f=f(\phi,n)$ with the corresponding operator $\hat{f}$ defined by (\ref{eq1.5}) 
 the expectation value $\braket{\hat{f}}$ in a state $\hat{\varrho}$ is 
  \begin{equation}\label{eq1.9}
   \braket{\hat{f}} = \mathrm{Tr} \left\{ \hat{\varrho} \hat{f} \right\}= \sum_{n=0}^{\infty} \int_{-\pi}^{\pi} f(\phi,n) 
   \varrho_W(\phi,n)
   \mathrm{d}\phi.
 \end{equation}
 Finally, the marginal distributions read 
  \begin{subequations}
    \begin{align}
       \varrho(n) &:=  \int_{-\pi}^{\pi}     \varrho_W(\phi,n)   \mathrm{d}\phi =    \braket{n| \hat{\varrho} 
       |n}\label{eq1.10a},
       \\
          \varrho(\phi ) &:= \sum_{n=0}^{\infty} \varrho_W(\phi,n)  =   \braket{\phi| \hat{\varrho} 
          |\phi}\label{eq1.10b}.
    \end{align}
  \end{subequations}
 In particular for any function $f=f(\phi)$ Eqs (\ref{eq1.9}) and (\ref{eq1.10b}) 
 give
 \begin{equation}\label{eq1.11}
   \braket{ \hat{f}} = \int_{-\pi}^{\pi}    f(\phi) \varrho(\phi)   \mathrm{d}\phi 
   =: \braket{f(\phi)}
 \end{equation}
 with $\hat{f}$ given by (\ref{eq1.8}). As it has been pointed out by many 
 authors \cite{2,3,4,5,6} the formula (\ref{eq1.11}) gives the expectation value 
 for the quantum phase-function $f(\phi)$ equal to the one calculated within the 
 Pegg-Barnett approach to quantum phase \cite{7,8,9}.
 Consequently, our choice for a number-phase Wigner function (\ref{eq1.4}) seems 
 to be appropriate when the Pegg-Barnett formulation is considered.
 A crucial question one should answer is whether the given number-phase Wigner 
 function $\varrho_{W}(\phi,n)$ defines uniquely the respective density operator 
 $\hat{\varrho}$. In Ref. \cite{2} we have suggested that this is indeed the 
 case (see Corrigendum in \cite{2}) but we were not able to give any explicit 
 formula therein. We fill this gap in Section \ref{sec2} of the present paper. 
 Another important issue which is explored in Section \ref{sec3} is the relation 
 between $\varrho_{W}$ and other well known in quantum optics quasiprobability 
 distributions such as the \textit{Glauber-Sudarshan distribution 
 $\mathcal{P}(\alpha)$},
 the \textit{Husimi function $\mathcal{Q}(\alpha)$} or the \textit{Wigner function 
 $\mathcal{W}(\alpha)$}.
 We find the formula which enables one to obtain the number-phase Wigner 
 function $\varrho_W(\phi,n)$ from those distributions and, in general from the 
 \textit{Cahill-Glauber function $\mathcal{W}^{(s)}(\alpha),\, -1\leq s \leq 
 1$}. 
 We hope that the results of the present work show that the function $\varrho_W(\phi,n)$ 
 can be considered as some useful quasiprobability distribution depending on 
 the photon number $n$ and on the phase $\phi$ (see Section \ref{sec4} where some conclusions are 
 given).
 
  \section{From the number-phase Wigner function $\varrho_W(\phi,n)$ to the respective density operator $\hat{\varrho}$} \label{sec2}

  First, we recall some properties of the celebrated Susskind-Glogower (SG) 
  phase operators $\widehat{e^{i\phi}}$ and $\widehat{e^{-i\phi}}$ 
  \cite{10,11,4,5,12,13}. They can be defined as 
  \begin{equation}\label{eq2.1}
    \widehat{e^{i\phi}} =
     \int_{-\pi}^{\pi} e^{i\phi} 
    \ket{\phi}\bra{\phi}\mathrm{d}\phi =   \sum_{n=0}^{\infty} \ket{n}\bra{n+1}
  \end{equation}
 and
   \begin{equation}\label{eq2.2}
    \widehat{e^{-i\phi}} =
     \int_{-\pi}^{\pi} e^{-i\phi} 
    \ket{\phi}\bra{\phi}\mathrm{d}\phi =   \sum_{n=0}^{\infty} \ket{n+1}\bra{n}= 
    \left( \widehat{e^{i\phi}}  \right)^{\dagger}.
  \end{equation}
  One can easily show that the annihilation $\hat{a}$ and creation $\hat{a}^{\dagger}$ 
  operators can be expressed in terms of the SG operators as follows 
  \begin{equation}\label{eq2.3}
    \hat{a} = \sqrt{\hat{n}+1}\,  \widehat{e^{i\phi}} =   \widehat{e^{i\phi}} \sqrt{\hat{n}}   
  \end{equation}
  and
    \begin{equation}\label{eq2.4}
    \hat{a}^{\dagger} = \widehat{e^{-i\phi}} \sqrt{\hat{n}+1} =   \sqrt{\hat{n}}   \,\widehat{e^{-i\phi}}. 
  \end{equation}
  For any operator depending on $\hat{n}$ $f=f(\hat{n})$ one gets
  \begin{equation}\label{eq2.5}
    \begin{split}
       \widehat{e^{i\phi}} f(\hat{n}) &= f(\hat{n}+1) \widehat{e^{i\phi}}\\
       \widehat{e^{-i\phi}} f(\hat{n}+1) &= f(\hat{n}) \widehat{e^{-i\phi}}
    \end{split}
  \end{equation}
   (compare with (\ref{eq2.3}) ad (\ref{eq2.4})). From (\ref{eq2.1}) and (\ref{eq2.2}) 
    we quickly have
    \begin{equation}\label{eq2.6}
       \widehat{e^{i\phi}} \ket{n} = \ket{n-1}, \quad  
       \widehat{e^{-i\phi}}\ket{n}=\ket{n+1}.
    \end{equation}
So $ \widehat{e^{i\phi}}$ and $ \widehat{e^{-i\phi}}$    are certain lowering 
and raising operators, respectively. Finally, employing (\ref{eq2.1}), (\ref{eq2.2}) 
and the general definition (\ref{eq1.8}) one finds the relation
\begin{equation}\label{eq2.7}
  \begin{split}
    \left(\widehat{e^{i\phi}}  \right)^k  \left(\widehat{e^{-i\phi}}  \right)^m &=
  \begin{cases}
  \left(\widehat{e^{i\phi}}  \right)^{k-m},   & \quad k \geq m\\
 \left(\widehat{e^{-i\phi}}  \right)^{m-k},  & \quad  k<m\\
  \end{cases}\\
  &= \int_{-\pi}^{\pi} e^{i(k-m)\phi} \ket{\phi}\bra{\phi} \mathrm{d}\phi = \widehat{e^{i(k-m)\phi}} 
  .
  \end{split}
\end{equation}
    Using (\ref{eq2.3}), (\ref{eq2.4}), (\ref{eq2.5}) and (\ref{eq2.7}), after 
    performing straightforward calculations we get
\begin{equation}\label{eq2.8}
  \begin{split}
    \hat{a}^m &= \sqrt{ (\hat{n}+1)(\hat{n}+2)\dots(\hat{n}+m) }\left(\widehat{e^{i\phi}}  
    \right)^m\\
        \left( \hat{a}^{\dagger}\right) ^m &= \left(\widehat{e^{-i\phi}}  
    \right)^m\sqrt{ (\hat{n}+1)(\hat{n}+2)\dots(\hat{n}+m) }\\
     \hat{a}^m   \left( \hat{a}^{\dagger}\right) ^m 
     &=(\hat{n}+1)(\hat{n}+2)\dots(\hat{n}+m)\\
     \hat{a}^m   \left( \hat{a}^{\dagger}\right) ^k &= 
     (\hat{n}+m-k+1)(\hat{n}+m-k+2)\dots(\hat{n}+m)\cdot \\
     &\cdot\sqrt{(\hat{n}+1)(\hat{n}+2)\dots(\hat{n}+m-k)}  \left(\widehat{e^{i\phi}}  
     \right)^{m-k}, \quad m>k\\
      \hat{a}^k   \left( \hat{a}^{\dagger}\right) ^m &=   \left(\widehat{e^{-i\phi}}  \right)^{m-k}
     (\hat{n}+m-k+1)(\hat{n}+m-k+2)\dots(\hat{n}+m)\cdot \\
     &\cdot\sqrt{(\hat{n}+1)(\hat{n}+2)\dots(\hat{n}+m-k)},\quad m>k.
  \end{split}
\end{equation} 
We express the density operator $\hat{\varrho}$ in the form 
\begin{equation}\label{eq2.9}
  \hat{\varrho} = \sum_{m,k=0}^{\infty} A_{mk}\hat{a}^m\left( \hat{a}^{\dagger}\right)^k
\end{equation}
i.e. as a series of operators constructed from $\hat{a}$ and $\hat{a}^{\dagger}$ in 
anti-normal ordering.
Inserting (\ref{eq2.8}) into (\ref{eq2.9}) and using the last equality of (\ref{eq2.7}) 
one obtains
\begin{equation}\label{eq2.10}
  \hat{\varrho} = \sum_{m=0}^{\infty} \left\{ \varrho_m(\hat{n})\widehat{e^{im\phi}} +  \widehat{e^{-im\phi}}\left(\varrho_m(\hat{n})\right)^{\dagger}  \right\}
\end{equation}
where $\varrho_m(\hat{n}),\, m=0,1,\dots$, are some operators, depending on 
$\hat{n}$.
Substituting (\ref{eq2.10}) into (\ref{eq1.4}) we get the number-phase Wigner 
function $\varrho_{W}(\phi,n)$ in the form 
\begin{equation}\label{eq2.11}
 \begin{split}
   \varrho_W(\phi,n) &= \frac{1}{2\pi} \mathrm{Re} 
    \Bigg\{ 
     \sum_{m=0}^{\infty} \varrho_m(n-m) e^{im\phi} +
     \sum_{m=0}^{\infty} \varrho_m^*(n) e^{-im\phi}
   \Bigg\}\\
   &= \frac{1}{4\pi}
       \Bigg\{ 
     \sum_{m=0}^{n}\left[   \left(\varrho_m(n-m) +\varrho_m(n) \right)   e^{im\phi} + 
     \left(\varrho_m^*(n-m) + \varrho_m^*(n) \right)e^{-im\phi}  \right]\\
     &+
     \sum_{m=n+1}^{\infty}   \left(  \varrho_m(n) e^{im\phi}+\varrho_m^*(n) e^{-im\phi}  \right) 
   \Bigg\}.
 \end{split}
\end{equation}
From (\ref{eq2.11}) one immediately has 
  \begin{subequations}
    \begin{align}
       \varrho_m(n) &= 2 \int_{-\pi}^{\pi}   \varrho_W(\phi,n)e^{-im\phi}  \mathrm{d}\phi,\quad m\geq n+1   \label{eq2.12a},
       \\
       \varrho_m(n-m) +       \varrho_m(n)&= 2 \int_{-\pi}^{\pi}   \varrho_W(\phi,n)e^{-im\phi}  \mathrm{d}\phi,\quad 1 \leq m \leq n\label{eq2.12b},
\\
       \varrho_0(n) +       \varrho_0^*(n)&= \int_{-\pi}^{\pi}   \varrho_W(\phi,n)  \mathrm{d}\phi\label{eq2.12c}
       \end{align}
  \end{subequations}
and we quickly conclude that to have all ingredients determining $\hat{\varrho}$ 
from a given $\varrho_W(\phi,n)$ we need $\varrho_m(n)$ for $1 \leq m \leq n$ (see 
(\ref{eq2.12b})).
Simple but rather tedious inductive analysis of (\ref{eq2.12a}) and (\ref{eq2.12b}) 
leads to the result
\begin{equation}\label{eq2.13}
  \varrho_m(n) = 2 \int_{-\pi}^{\pi} 
\left( \sum_{l=0}^{\left[ \frac{n}{m}\right]}  (-1)^l \varrho_W(\phi,n-lm) \right)
   e^{-im\phi}
  \mathrm{d}\phi, \quad1\leq m\leq n.
\end{equation}
Observe that the formula (\ref{eq2.13}) holds true also for $m\geq n+1$ as  
if $m\geq n+1$ then Eq. (\ref{eq2.13}) gives exactly (\ref{eq2.12a}). Gathering, we arrive 
at the conclusion
\begin{equation}\label{eq2.14}
  \hat{\varrho} = \varrho_0(\hat{n}) + \left( \varrho_0(\hat{n})\right)^{\dagger}  + 
  \sum_{m=1}^{\infty} \left\{   \varrho_m(\hat{n})  \widehat{e^{im\phi}}  + \widehat{e^{-im\phi}} \left(  \varrho_m(\hat{n})\right)^{\dagger}\right\}
\end{equation}
where 
 \begin{subequations}
    \begin{align}
        &\varrho_0(\hat{n}) + \left( \varrho_0(\hat{n})\right)^{\dagger}  = \sum_{n=0}^{\infty}
  \left\{   \int_{-\pi}^{\pi} \varrho_W(\phi,n)\mathrm{d}\phi \ket{n}\bra{n}  
  \right\}\label{eq2.15a}\\
      &\varrho_m(\hat{n})  = \sum_{n=0}^{\infty}
  \left\{  2 \int_{-\pi}^{\pi}    
  \left( \sum_{l=0}^{\left[ \frac{n}{m}\right]}  (-1)^l \varrho_W(\phi,n-lm) 
  \right)e^{-im\phi}
  \mathrm{d}\phi \ket{n}\bra{n}  
  \right\},\quad m \geq 1.\label{eq2.15b}
    \end{align}
 \end{subequations}
    The formulae (\ref{eq2.14}), (\ref{eq2.15a}) and (\ref{eq2.15b}) 
   show that  the density operator $\hat{\varrho}$ is \textit{defined uniquely} by the 
   respective number-phase Wigner function $\varrho_W(\phi,n)$.
   From  (\ref{eq2.14}) with  (\ref{eq2.15a}),  (\ref{eq2.15b}) and  (\ref{eq2.6}) 
   one can derive the following useful relations 
   \begin{equation}\label{eq2.16}
     \begin{split}
       \varrho_m(n) &= \braket{n|\varrho_m(\hat{n})|n} = \braket{n |
       \hat{\varrho}|n+m}, \quad m\geq1,\\
       \varrho_0(n)+\varrho_o^*(n) &= \braket{n |\varrho_0(\hat{n}) + \left( \varrho_0(\hat{n})\right)^+ |n  
       }= \braket{n|\hat{\varrho}|n}.
     \end{split}
   \end{equation}
 
\begin{figure}[h]
\centering
\includegraphics{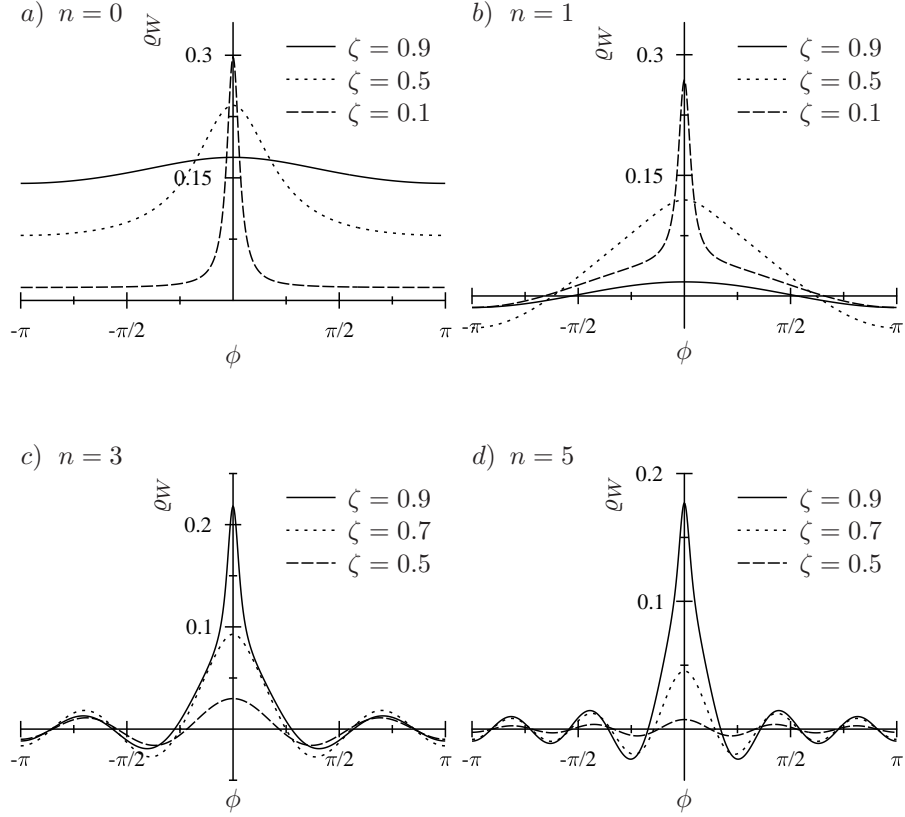}
\caption{\label{fig1} Plots of $\varrho_W(\phi,n)$ with $\varphi =0$ for $a)$: $n=0$, $b)$: $n=1$,
$c)$: $n=3$ and $d)$: $n=5$        }
\end{figure}

 \textbf{Example:}
 As a simple example consider the \textit{coherent phase states} \cite{5,14} 
 represented by the kets 
 \begin{equation}\label{eq2.17}
   \ket{\zeta} = (1-|\zeta|^2)^{\frac{1}{2}} \sum_{n=0}^{\infty} \zeta^n 
   \ket{n}, \quad |\zeta|<1
 \end{equation}
 which are the  normalised  eigenvectors of the lowering operator  $\widehat{e^{i\phi}}$
 \begin{equation}\label{2.18}
   \widehat{e^{i\phi}} \ket{\zeta} = \zeta\ket{\zeta}, \quad 
   \braket{\zeta|\zeta}=1.
 \end{equation}
 The respective density operator $\hat{\varrho}$ reads
 \begin{equation}\label{eq2.19}
   \hat{\varrho} = \ket{\zeta}\bra{\zeta} = (1-|\zeta|^2)  \sum_{k,l=0}^{\infty}  
   |\zeta|^{k+l} e^{i(k-l)\varphi}\ket{k}\bra{l}
 \end{equation}
 where $\zeta=|\zeta|e^{i\varphi}$.
 Inserting (\ref{eq2.19}) into (\ref{eq1.4}) one easily gets the corresponding  
 number-phase  Wigner function
 \begin{equation}\label{eq2.20}
\begin{split}
  \varrho_W(\phi,n) &= \frac{(1-|\zeta|^2)|\zeta|^n}{2\pi} \sum_{k=0}^{\infty}   |\zeta|^k \cos{\left[ 
  (n-k)(\phi-\varphi)\right]}\\
  &= \frac{(1-|\zeta|^2)|\zeta|^n}{2\pi}
  \Bigg\{ 
  \cos{\left[n(\phi-\varphi)\right]} \sum_{k=0}^{\infty}   |\zeta|^k \cos{\left[ 
  k(\phi-\varphi)\right]} \\
  &+ \sin{\left[n(\phi-\varphi)\right]} \sum_{k=0}^{\infty}   |\zeta|^k \sin{\left[ 
  k(\phi-\varphi)\right]}
  \Bigg\}
\end{split} 
 \end{equation}
 [Compare with the number-phase Wigner function for the coherent state $\ket{\alpha}$ (see Eq. (6.6) of 
 \cite{2})].
 Figure \ref{fig1} shows exemplary plots of the number-phase Wigner function for the coherent phase 
 state with $\varphi=0$ and various values of $n$ and $|\zeta|$. 
 Note that for $n=0$ the $\varrho_W(\phi ,n)$ is positive for every value of $\zeta$ 
(see Figure \ref{fig1} $a$). 

 Substituting (\ref{eq2.19}) into (\ref{eq2.16}) we find
 \begin{equation}\label{eq2.21}
   \begin{split}
     \varrho_m(n) &= (1-|\zeta|^2) |\zeta|^{2n+m} e^{-im\varphi}, \quad m\geq 1 
     \\
     \varrho_0(n)+\varrho_0(n)^* &= (1-|\zeta|^2) |\zeta|^{2n}.
   \end{split}
 \end{equation}
 Hence
  \begin{equation}\label{eq2.22}
   \begin{split}
     \varrho_m(\hat{n}) &=   \sum_{n=0}^{\infty} \varrho_m(n)\ket{n}\bra{n}\\
     &=(1-|\zeta|^2) |\zeta|^{m} e^{-im\varphi} \sum_{n=0}^{\infty} |\zeta|^{2n} \ket{n}\bra{n}, \quad m\geq 1 
     \\
     \varrho_0(\hat{n})+\left(\varrho_0(\hat{n})\right)^{\dagger} &= (1-|\zeta|^2)  \sum_{n=0}^{\infty} 
     |\zeta|^{2n} \ket{n}\bra{n}.
   \end{split}
 \end{equation}
 It is an easy matter to show that inserting (\ref{eq2.22}) into (\ref{eq2.14}) 
 and employing (\ref{eq2.6}) one gets the density operator (\ref{eq2.19})


 \section{Relation between $\varrho_W(\varphi,n)$ and other quasiprobability distributions.}\label{sec3}
 
 We begin our considerations with searching for a relation between $\varrho_W(\varphi,n)$ 
 and the celebrated Glauber-Sudarshan function $\mathcal{P}(\alpha)$ defined by 
 \cite{15,16,17,18}
 \begin{equation}\label{eq3.1}
   \hat{\varrho} = \int \mathcal{P}(\alpha) \ket{\alpha}\bra{\alpha} \mathrm{d}^2 
   \alpha,\quad \alpha\in \mathbb{C}, \mathrm{d}^2 
   \alpha = \mathrm{d\,Re} \alpha \cdot \mathrm{d\,Im} \alpha
 \end{equation}
 where $\ket{\alpha}$ stands for the normalised ket vector of the coherent state 
 \begin{equation}\label{eq3.2}
   \begin{split}
   \hat{a} \ket{\alpha } &= \alpha \ket{\alpha}, \\ 
   \ket{\alpha}&= e^{-\frac{1}{2} |\alpha|^2}
    \sum_{n=0}^{\infty}  \frac{\alpha^n}{\sqrt{n!}}\ket{n},\\
    \braket{\alpha|\alpha} &=1, \quad \braket{\alpha|\beta} =e^{-\frac{1}{2} (|\alpha|^2 + |\beta|^2 - 
    \alpha^*\beta)},\quad \alpha,\beta \in \mathbb{C}.
   \end{split}
 \end{equation}
 From the last equation of (\ref{eq3.2}) one quickly concludes that two 
 different coherent states are non orthogonal. However, as is well known the set 
$\{ \ket{\alpha}\}_{\alpha \in \mathbb{C}}$
 gives a resolution of the identity operator
 \begin{equation}\label{eq3.3}
   \frac{1}{\pi} \int \ket{\alpha}\bra{\alpha} \mathrm{d}^2\alpha = \hat{1}.
 \end{equation}
 Assuming $\hat{\varrho}$ in the form of (\ref{eq2.9}) and using (\ref{eq3.3}) 
 and then the first formula of (\ref{eq3.2}) one gets
 \begin{equation}\label{eq3.4}
   \begin{split}
     \hat{\varrho} &=\frac{1}{\pi} \int \sum_{m,k=0}^{\infty} A_{mk}  \hat{a}^m \ket{\alpha}\bra{\alpha} 
     (\hat{a}^{\dagger})^k \mathrm{d}^2 \alpha \\
     &= \int \left( \frac{1}{\pi}  \sum_{m,k=0}^{\infty}  A_{mk} \alpha^m (\alpha^*)^k \right) 
     \ket{\alpha}\bra{\alpha} \mathrm{d}^2 \alpha.
        \end{split}
 \end{equation}
 Comparing (\ref{eq3.1})  and (\ref{eq3.4}) we immediately have 
 \begin{equation}\label{eq3.5}
   \mathcal{P}(\alpha) = \frac{1}{\pi} \sum_{m,k=0}^{\infty}  A_{mk} \alpha^m 
   (\alpha^*)^k.
 \end{equation}
 To express $\varrho_w(\phi,n)$ by $\mathcal{P}(\alpha)$ we insert (\ref{eq3.1})  
 into (\ref{eq1.4}) and we obtain
 \begin{equation}\label{eq3.6}
   \begin{split}
     \varrho_W(\phi,n) &= \frac{1}{2\pi} \int \mathcal{P}(\alpha) \mathrm{Tr} \left\{ \ket{\alpha}\bra{\alpha}
     \hat{\Omega}(\phi,n) \right\}\mathrm{d}^2 \alpha \\
     &= \mathrm{Re} \left\{   \int \mathcal{P}(\alpha) 
     \braket{\phi |\alpha}    \braket{\alpha|n} \braket{n|\phi} \mathrm{d}^2 \alpha
   \right\}.
   \end{split}
 \end{equation}
 Employing (\ref{eq1.3}) and (\ref{eq3.2}), and performing simple calculations 
 one  gets
 \begin{equation}\label{eq3.7}
     \varrho_W(\phi,n) = \frac{1}{2\pi\sqrt{n!}} \int \mathcal{P}(\alpha)
     |\alpha|^n e^{-|\alpha|^2} 
     \sum_{k=0}^{\infty} \frac{|\alpha|^k}{\sqrt{k!}}  \cos\left[(n-k)(\phi-\gamma) \right] 
     \mathrm{d}^2 \alpha
 \end{equation}
 where $\alpha= |\alpha| e^{i\gamma}$, $\mathrm{d}^2 \alpha = |\alpha| \,\mathrm{d}|\alpha|\cdot \mathrm{d} 
 \gamma$.
 
 Now we are going to study the relation between $\varrho_W(\phi,n)$ and the 
 \textit{Cahill-Glauber function $\mathcal{W}^{(s)}(\alpha)$} defined as \cite{19,18}
 \begin{equation}\label{eq3.8}
   \mathcal{W}^{(s)}(\alpha) = \frac{1}{\pi} \mathrm{Tr} \left\{ \hat{\varrho} \hat{T}^{(s)}(\alpha) 
   \right\},\quad s\in [-1,1],\quad \alpha \in \mathbb{C}
 \end{equation}
 where the operator $\hat{T}^{(s)}(\alpha)$ is given by
 \begin{equation}\label{eq3.9}
   \hat{T}^{(s)}(\alpha) = \frac{1}{\pi} \int 
   e^{\alpha \xi^* - \alpha^* \xi + \frac{s}{2}|\xi|^2}\hat{D}(\xi) 
   \mathrm{d}^2 \xi, \quad \xi \in \mathbb{C},\quad \mathrm{d}^2\xi = \mathrm{d} 
   \mathrm{Re}\xi \cdot  \mathrm{d} 
   \mathrm{Im}\xi
 \end{equation}
 with $\hat{D}(\xi)$ standing for the \textit{displacement operator} 
 \begin{equation}\label{eq3.10}
   \hat{D}(\xi) = e^{\xi \hat{a}^{\dagger} -\xi^* \hat{a}} = e^{-\frac{1}{2} 
   |\xi|^2}e^{\xi \hat{a}^{\dagger}} e^{-\xi^* \hat{a}}=
    e^{\frac{1}{2} 
   |\xi|^2}e^{-\xi^* \hat{a}} e^{\xi \hat{a}^{\dagger}}.
 \end{equation}
 Note that for $s=-1$, the Cahill-Glauber function $\mathcal{W}^{(-1)}(\alpha)$ 
 is usually denoted by ${Q}(\alpha)$ and it is well known in quantum optics 
 as the \textit{Husimi function};
 for $s=0$, $\mathcal{W}^{(0)}(\alpha)$ is the famous \textit{Wigner function} 
 and for $s=1$ we get $\mathcal{W}^{(1)}= \mathcal{P}(\alpha)$ 
 \cite{17,18,19,20,21}.
 Given $\mathcal{W}^{(s)}(\alpha)$, $-1\leq s \leq 1$, one can find the 
 respective density operator $\hat{\varrho}$ from the following formula (see for instance \cite{18,19})
 \begin{equation}\label{eq3.11}
   \hat{\varrho} = \int \mathcal{W}^{(s)}(\alpha) 
   \hat{T}^{(-s)}(\alpha)\mathrm{d}^2\alpha.
 \end{equation}
 Substituting (\ref{eq3.11}) into (\ref{eq1.4})   we get
 \begin{equation}\label{eq3.12}
   \begin{split}
     \varrho_W(\phi,n) &= \frac{1}{2\pi} \int \mathcal{W}^{(s)}(\alpha) 
     \mathrm{Tr} \left\{      \hat{T}^{(-s)}(\alpha) \hat{\Omega}(\phi,n)  \right\}\mathrm{d}^2  \alpha
   \\
     &= \frac{1}{4\pi} \int \mathcal{W}^{(s)}(\alpha) 
     \sum_{k=0}^{\infty}
     \bigg\{ 
     \braket{k | \hat{T}^{(-s)}(\alpha)  |n} e^{i(n-k)\phi} \\&+
               \braket{n| \hat{T}^{(-s)}(\alpha)  |k} e^{-i(n-k)\phi}
     \bigg\}
     \mathrm{d}^2 
     \alpha.
   \end{split}
 \end{equation}
 The matrix elements of $  \hat{T}^{(-s)}(\alpha) $ are given by (see e.g. \cite{18,19})
 \begin{equation}\label{eq3.13}
   \begin{split}
          \braket{j | \hat{T}^{(-s)}(\alpha)  |k}&= \braket{k| \hat{T}^{(-s)}(\alpha)  
          |j}^* =  \sqrt{\frac{j!}{k!} } \left( \frac{2}{s+1}\right)^{k-j+1} \left( \frac{s-1}{s+1}\right)^{j}
          \left( \alpha^*\right)^{k-j}\\
          &\cdot \exp{\left( - \frac{2|\alpha|^2}{s+1}\right)} L_j^{k-j} \left( \frac{4 |\alpha|^2}{1-s^2}\right)
   \end{split}
 \end{equation}
 where $L_j^{k-j}$ stands for the associated Laguerre polynomial.
 Inserting (\ref{eq3.13}) into (\ref{eq3.12}), after some simple manipulations 
 one arrives at the formula 
 \begin{equation}\label{eq3.14}
 \begin{split}
      \varrho_W(\phi,n) &= \frac{\sqrt{n!}(s-1)^n} {2^n (s+1)\pi}
      \int_0^{\infty} \mathrm{d} |\alpha| \int_{-\pi}^{\pi} \mathrm{d} \gamma \,
      \mathcal{W}^{(s)}(|\alpha| e^{i\gamma}) |\alpha'|^{-n+1}
   \exp{\left( -\frac{2|\alpha|^2}{s+1}\right)} \\    &\cdot \sum_{k=0}^{\infty}
       \frac{1}{\sqrt{k!} }
             \left( \frac{2|\alpha|}{s+1}\right)^k 
      L_n^{k-n} \left( \frac{4|\alpha|^2}{1-s^2}\right)
      \cos{\left[ (k-n)(\phi-\gamma)\right]}
 \end{split}
 \end{equation}
 where we put $\alpha=|\alpha|e^{i\gamma}$.
 
 Although the relation (\ref{eq3.14}) is rather involved it shows that, in 
 principle, given the Cahill-Glauber function  $\mathcal{W}^{(s)}(\alpha)$, $-1 \leq s \leq 
 1$, we can find the respective number-phase Wigner function 
 $\varrho_W(\phi,n)$.  Of course, since $\varrho_W(\phi,n)$ defines uniquely the 
 density operator $\hat{\varrho}$ (see section \ref{sec2}) and $\hat{\varrho}$ 
 defines $\mathcal{W}^{(s)}(\alpha)$ by (\ref{eq3.8}) one concludes that given $\varrho_W(\phi,n)$
 we can find the respective $\mathcal{W}^{(s)}(\alpha)$.
 However the relevant formula seems to be extremely involved.
 

 \section{Conclusions}\label{sec4}
 
 We conclude that the proposed number-phase Wigner function $\varrho_W(\phi,n)$ 
 can be used as a quasiprobability distribution in quantum optics.
 The natural question is what are the relations between $\varrho_W(\phi,n)$  
 and other number-phase Wigner functions introduced previously by several 
 authors \cite{12,23,24,25}. 
 Another interesting question is if our consideration of the number-phase Wigner 
 function can be carried over to the case of a quantum system with finite 
 dimensional space of states. If it can be done then the respective procedure 
 would provide us with an approach to the Weyl-Wigner-Moyal formalism in quantum 
 mechanics of discrete systems alternative to the one given by S.~Chaturvedi et 
 al \cite{26}.
 We are going to consider these problems very soon. 
 



\end{document}